\documentclass[traditabstract]{aa} 
\usepackage{txfonts}
\usepackage{graphicx}

\begin{document}

\title{Testing X-ray selection effects with four rich, yet X--ray--faint, galaxy clusters}
\titlerunning{Four galaxy cluster among the richest, yet the X-ray faintest}
\author{S. Andreon$^{1}$\thanks{E-mail:stefano.andreon@inaf.it}, A. Moretti$^{1}$}
\authorrunning{Andreon \& Moretti}
\institute{
$^1$ INAF--Osservatorio Astronomico di Brera, via Brera 28, 20121, Milano, Italy\\
}
\date{Accepted ... Received ..; in original form ..}
\abstract{
A robust understanding of selection effects in galaxy cluster studies is crucial for both astrophysical and cosmological applications. Examining clusters identified through different observational strategies, even in small numbers, helps to illuminate potential biases inherent to each method.
We selected four rich galaxy clusters in the Northern Hemisphere whose early Swift X-ray Telescope (XRT) observations indicated unusually low central X-ray emission, making them unlikely to be detected in X-ray surveys. Spectroscopic follow-up confirms that all four systems are genuine galaxy clusters, rather than projections of multiple clusters or groups along the line of sight. Their optical richness, estimated using one of the baseline Euclid cluster richness estimators, implies masses of $\log M_{200}/M_\odot \sim 14.6$ and independently confirms the absence of additional massive structures along the line of sight.
Deep XRT follow-up reveals highly disturbed X-ray morphologies: three clusters exhibit at least two distinct X-ray peaks, while the remaining cluster has an axis ratio exceeding 1.5. Spectroscopy shows that galaxies associated with different parts share the same redshift, demonstrating that these substructures are physically connected rather than chance projections.
These clusters display low central X-ray surface brightness and total X-ray luminosities suppressed by roughly one dex for their richness, making them undetectable in X-ray surveys as eROSITA.
We estimate $\sim$20\% as a lower limit for the poorly sampled population, albeit based on a small sample. Our results demonstrate that even rich clusters in the northern $z<0.3$ Universe can be missed by X-ray selection and that the X-ray variety captured by X-ray surveys underestimates the true cluster diversity.
}
\keywords{galaxies: clusters: general --- Galaxies: clusters: intracluster medium}
\maketitle

\section{Introduction}

Galaxy clusters can be identified in several ways: by searching for projected galaxy overdensities on the sky (e.g. Abell 1958; Zwicky 1961--1968), possibly exploiting galaxy colors and photometric redshifts to enhance the contrast against the field (e.g. Gladders \& Yee 2000; Andreon et al. 2003); by detecting extended emission in X-ray images tracing the hot intracluster medium (ICM; e.g. Gioia et al. 1990; Ebeling et al. 1998); by detecting a spatially localized distortion of the CMB spectrum due to the thermal Sunyaev--Zel'dovich (tSZ) effect (Sunyaev \& Zel'dovich 1972; Carlstrom et al. 2002), as in recent wide-area surveys (e.g. Reichardt et al. 2013; Planck Collaboration 2014); or by identifying concentrations of galaxies with concordant spectroscopic redshifts (e.g., Miller et al. 2005). Each technique probes a different physical observable of the underlying dark matter halo and is therefore characterized by specific advantages and limitations.

X-ray searches are said to provide a relatively clean selection of virialized systems, since the X-ray emissivity scales approximately with the square of the gas density, strongly suppressing projection effects (e.g. Bulbul et al. 2022). However, they require space-based observations with large effective area and sharp point spread function to detect clusters beyond the local Universe, as well as optical or spectroscopic data for confirmation and redshift determination. SZ surveys use a tracer that is nearly redshift independent and potentially less sensitive to the dynamical state of the system, albeit typically with lower angular resolution and potential contamination from astrophysical foregrounds. They also need optical or spectroscopic data for confirmation and redshift determination. Optical searches can efficiently cover very large areas from the ground and reach lower masses, but are often considered more susceptible to line-of-sight projections. However, in a controlled comparison in which projection is quantified within the same mass overdensity definition and samples are matched in number density, optical selection is not intrinsically more contaminated than X-ray or SZ selection (Andreon 2016). Owing to its superior redshift resolution, optical selection can in fact better separate structures along the line of sight, distinguishing systems with $\Delta z > 0.02$ (e.g. Gladders \& Yee 2000; Andreon \& Berge 2012) that are
instead blended in X-ray and SZ surveys.

A robust understanding of the selection effects induced by the cluster identification method is essential for both astrophysical and cosmological applications. Cluster abundances as a function of mass and redshift provide powerful cosmological constraints (e.g. Vikhlinin et al. 2009), but their interpretation relies on an accurate characterization of the survey selection function (Rosati et al. 1995) and of the amplitude of the population missed by the survey (e.g. Andreon \& Hurn 2013; Shirasaki et al. 2024; Andreon \& Radovich 2025). Similarly, scaling relations between observables (e.g. $L_{\rm X}$--$M$, $Y$--$M$, richness--$M$) are sensitive to selection biases (e.g. Pacaud et al. 2007; Viklinin et al. 2009; Mantz et al. 2010; Andreon et al. 2011, 2022 etc.). Inadequate modeling of these effects, or lack of knowledge about the size of the population below the observational threshold, propagates directly into biased cosmological parameters and into an incomplete understanding of the thermodynamical properties of the ICM.  

Examining clusters identified through different observational strategies, even in limited numbers, provides insight into the biases inherent to each method. This is particularly informative when the selected systems are among the richest or potentially most massive objects, such that their absence in another waveband is unlikely to be driven solely by stochastic scatter in multiwavelength scaling relations. Systems that are optically rich yet X--ray--faint for their richness offer a direct test of the connection between galaxy content and ICM properties, and of the assumptions underlying scaling relations calibrated on X-ray selected samples.

In this work, we select four rich clusters in the nearby Universe that appear faint in archival X-ray data and are therefore potential outliers of the richness--$L_{\rm X}$ relation defined by X-ray selected samples. These systems are candidates to be missed, or significantly down-weighted, in X-ray surveys. We combine new and archival spectroscopic, optical, and X-ray data to assess whether these objects are bona fide gravitationally bound clusters rather than chance superpositions, filamentary structures seen in projection, or systems substantially contaminated by foreground or background groups.

We assume $\Omega_M=0.3$, $\Omega_\Lambda=0.7$, and $H_0=70$ km s$^{-1}$ Mpc$^{-1}$. 
All logarithms are in base 10.

\begin{figure}
\centerline{\includegraphics[trim=105 90 30 20,clip,width=9truecm]{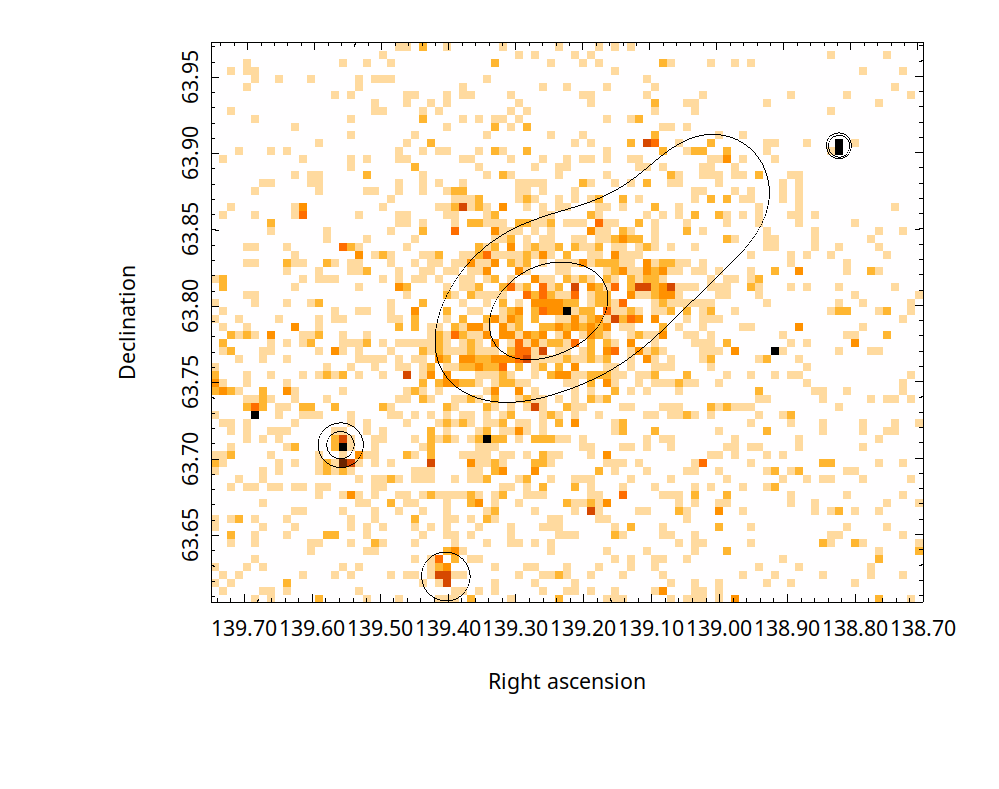}}
\caption[h]{Binned, exposure-uncorrected X-ray image of AM2 with superposed adaptively smoothed X-ray contours measured on the exposure-corrected image. The cluster is highly elongated, with an axis ratio of 1.5. The outer X-ray contour merges the emission of AM2 with that of a background cluster located to the NW. The apparent sharp discontinuity in surface brightness toward the NW is caused by a strong variation in exposure time across the mosaic, and is not a physical feature.}
\label{fig:AM2Xray}
\end{figure}

\begin{figure}
\centerline{\includegraphics[trim=40 30 40 20,clip,width=7truecm]{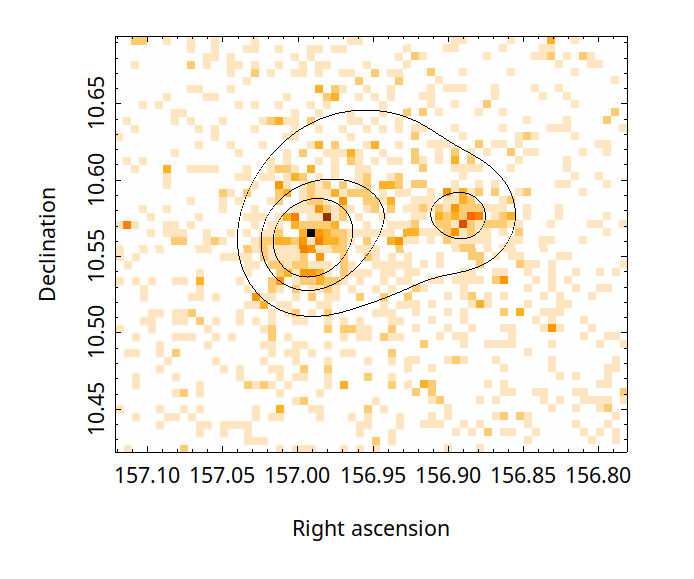}}
\caption[h]{As in Fig.~\ref{fig:AM2Xray}, but for AM7. The cluster is clearly bimodal, and spectroscopy confirms that both clumps belong to the same system rather than being two separate structures projected along nearby lines of sight.}
\label{fig:AM7Xray}
\end{figure}

\begin{figure}
\centerline{\includegraphics[trim=30 0 80 20,clip,width=9truecm]{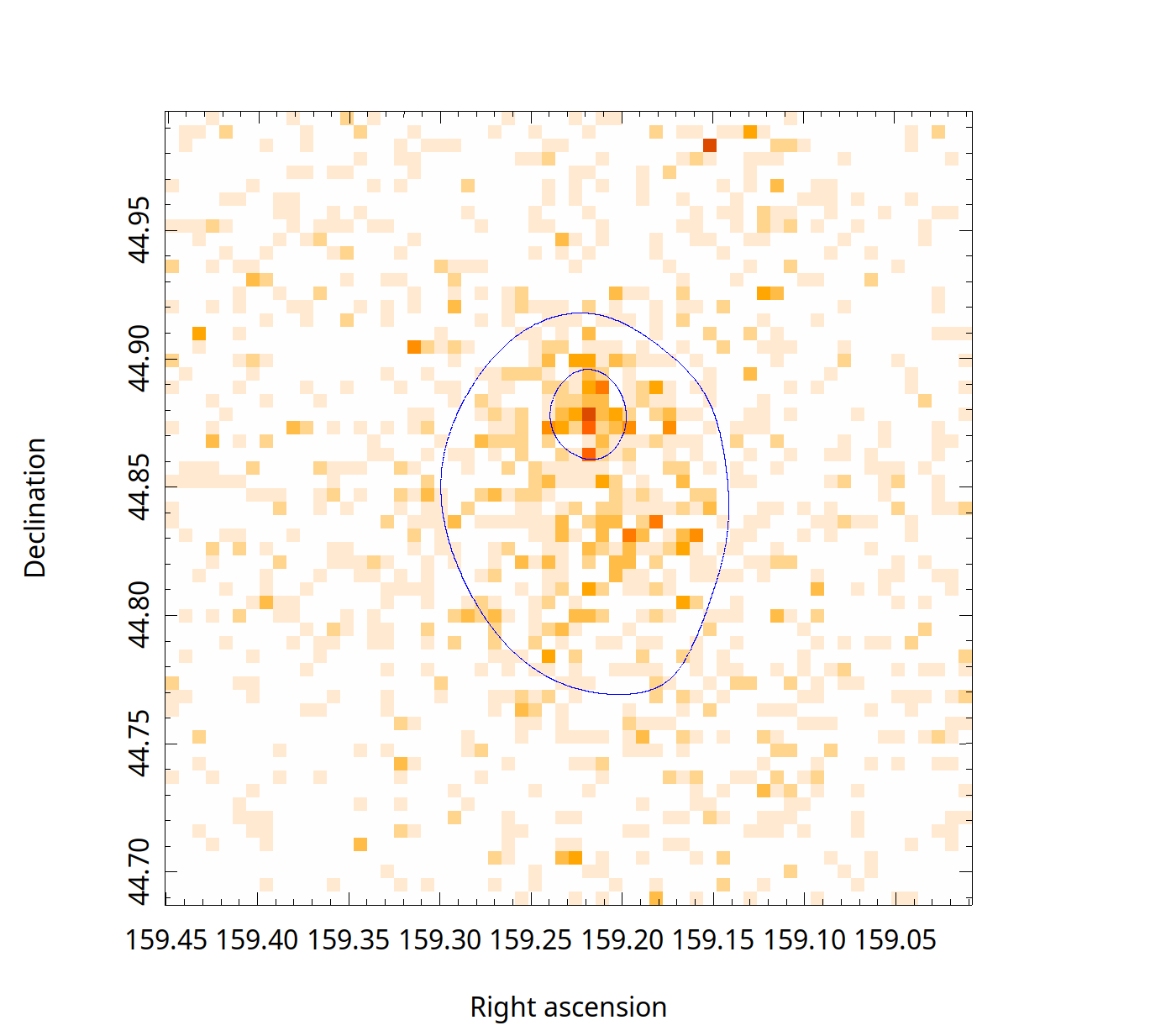}}
\caption[h]{As in Fig.~\ref{fig:AM2Xray}, but for AM62. The cluster is morphologically disturbed and asymmetric along the North--South direction. Spectroscopy confirms that this is a single system rather than two structures projected along nearby lines of sight.}
\label{fig:AM62Xray}
\end{figure}

\begin{figure}
\centerline{\includegraphics[trim=0 0 20 120,clip,width=7truecm]{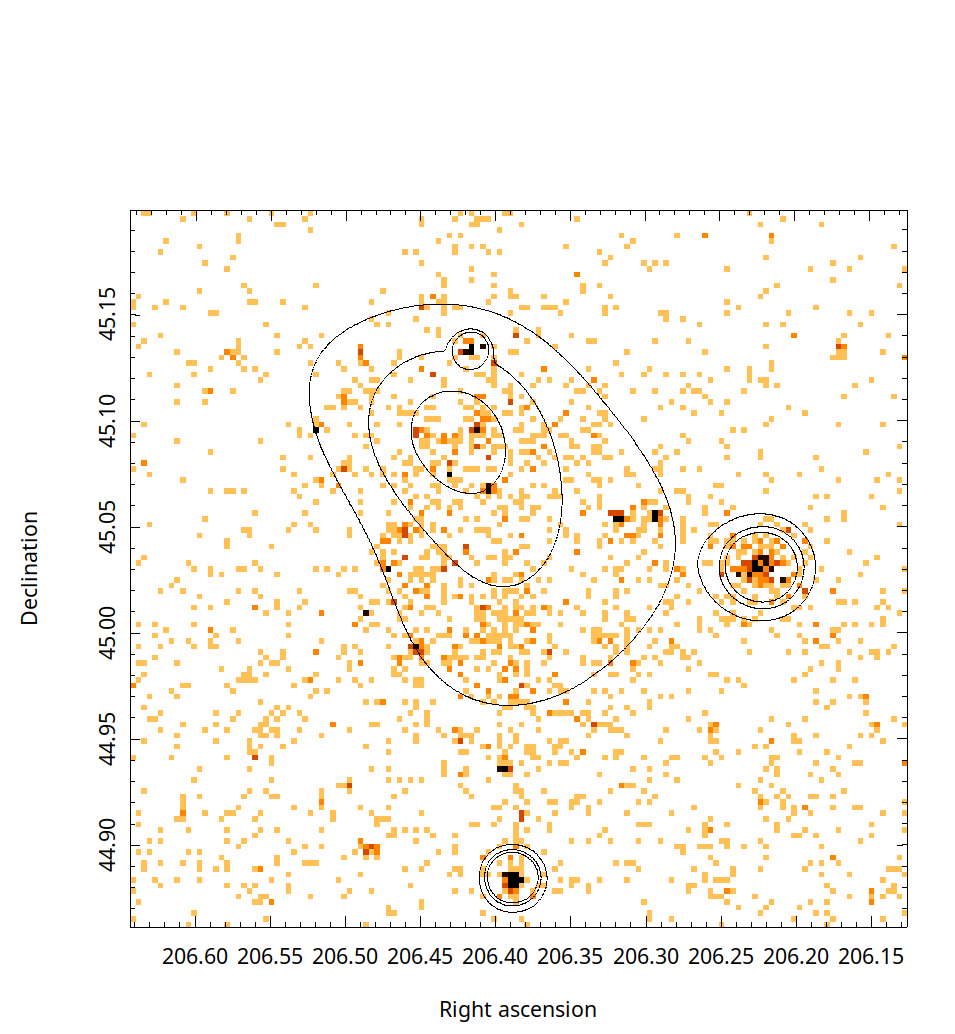}}
\caption[h]{As in Fig.~\ref{fig:AM2Xray}, but for AM72. The cluster is morphologically complex, with two clumps aligned along the North--South direction, also traced by the distribution of spectroscopic members in Fig.~\ref{fig:radecz}. Spectroscopy confirms that this is a single system rather than two structures projected along nearby lines of sight. The bright X-ray emission to the West corresponds to a background cluster. The extreme faintness of the cluster X-ray emission and the presence of several point sources prevent the contours from fully tracing the extended emission.}
\label{fig:AM72Xray}
\end{figure}

\begin{figure*}
\centerline{%
\includegraphics[trim=40 20 100 70,clip,width=7truecm]{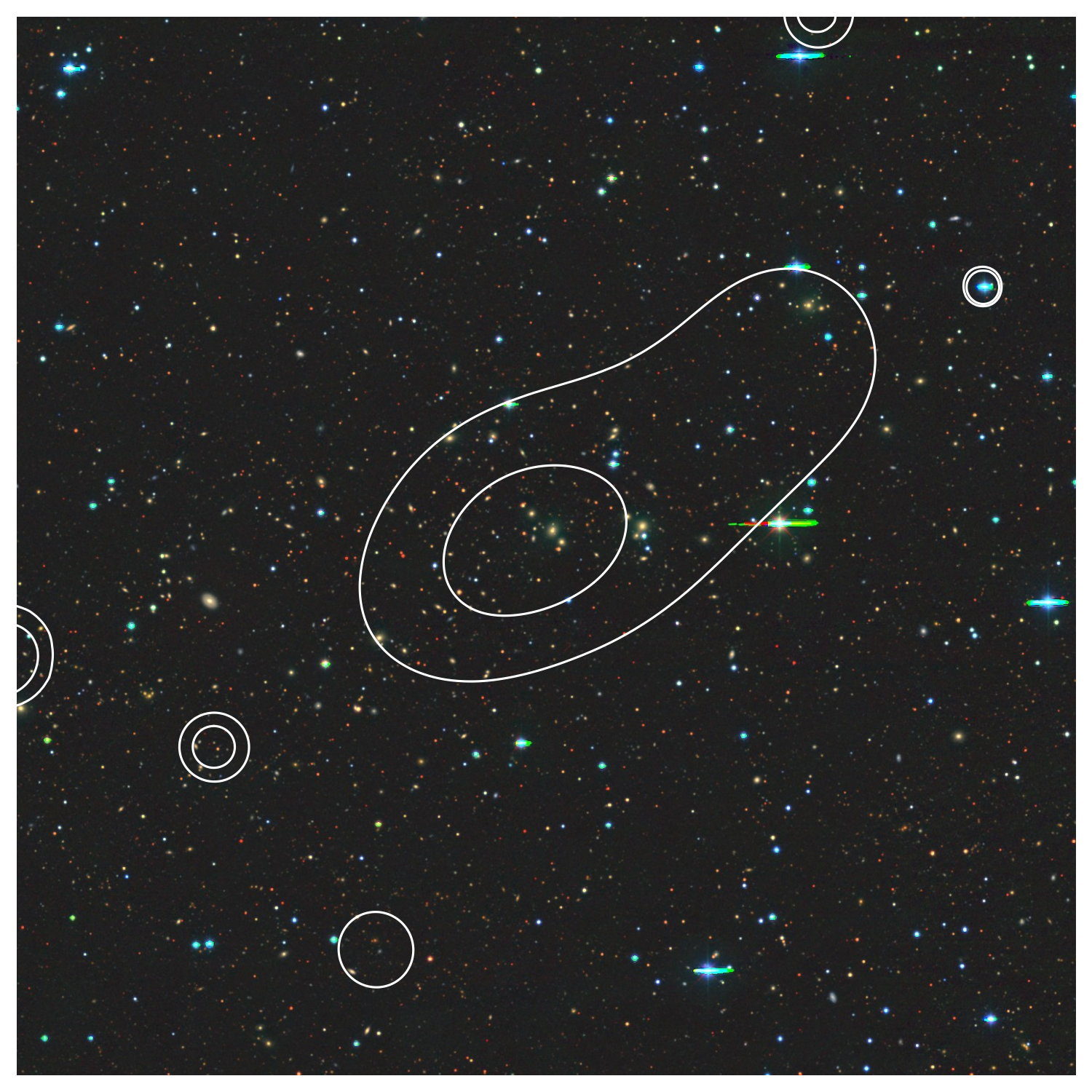}
\includegraphics[trim=40 20 100 70,clip,width=7truecm]{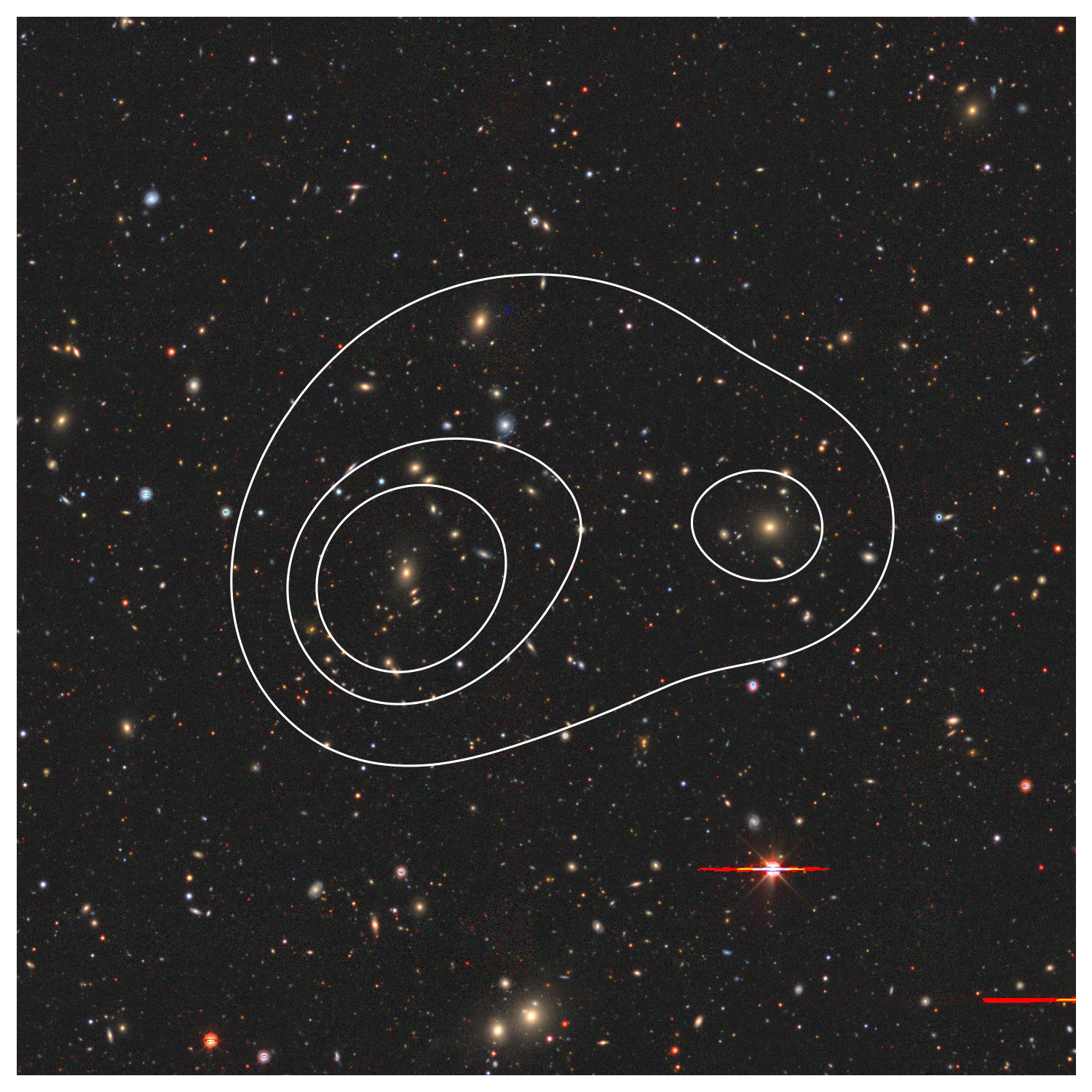}}%
\centerline{%
\includegraphics[trim=40 20 100 70,clip,width=7truecm]{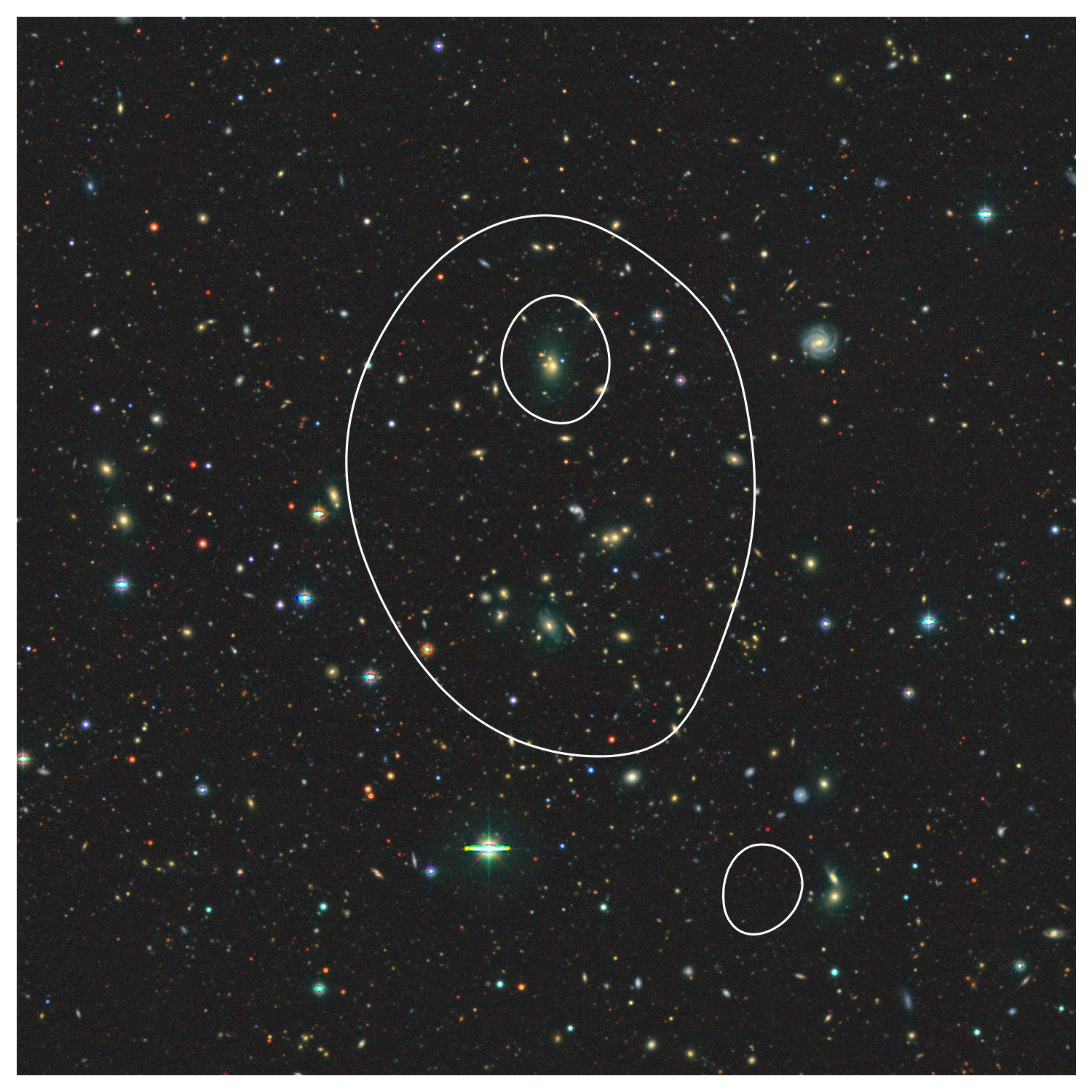}
\includegraphics[trim=40 20 100 70,clip,width=7truecm]{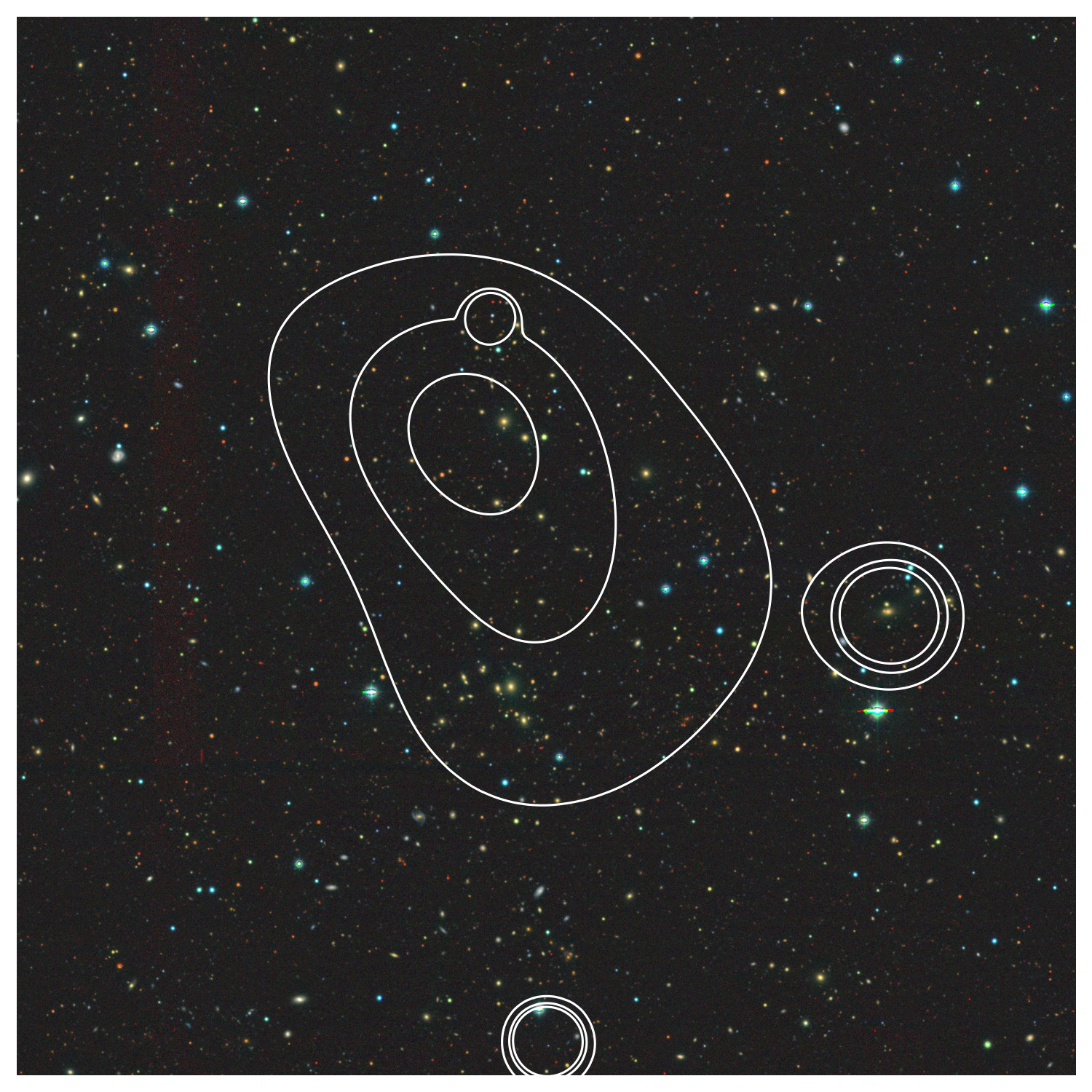}}
\caption[h]{True-color ($grz$) HSC images with overlaid X-ray contours derived from adaptively smoothed, exposure-corrected 0.5--2~keV images. North is up, East to the left. The field of view is a square of side 27, 17.5, 17.5, and 21.8 arcmin for AM2, AM7, AM62, and AM72 (left to right, top to bottom), respectively. The true-color images are taken from the Legacy Survey Viewer (https://www.legacysurvey.org/viewer).}
\label{fig:maps}
\end{figure*}

\setlength{\tabcolsep}{6pt}

\section{Sample selection, measurements, and results}

\subsection{Sample Selection}

We selected four nearby, rich clusters for dedicated X-ray follow-up, targeting systems that appeared to have low central X-ray surface brightness. The selection was based on clusters observed in early \textit{Swift}/XRT data, partly serendipitously and partly within our initial program aimed at observing all maxBCG clusters (Koester et al. 2007) with richness in the range $59 \le n \le 70$ and photometric redshift $0.1 < z_{\rm phot} < 0.3$ over a 7500\,deg$^2$ area. Results for the first observed systems were presented in Andreon \& Moretti (2011).

From this parent sample formed by 38 clusters with maxBCG richness larger than $n \gtrsim 50$ and XRT observations, we selected four clusters that appeared X--ray--faint within a 500\,kpc aperture compared to expectations based on their maxBCG richness, given the available data. For three of the four systems (all except AM62), the cluster position was located close to the edge of the \textit{Swift}/XRT field of view in the early observations. This reflects the substantial fraction of miscentered systems in the maxBCG catalog (Johnston et al. 2007; Andreon \& Moretti 2011), which affects the initial placement of X-ray pointings. 
Three additional clusters exhibit similar X-ray faintness and richness, but were not selected for observation in order to limit the requested telescope time.

\begin{figure*}
\centerline{\includegraphics[trim=0 0 0 0,clip,width=4.5truecm]{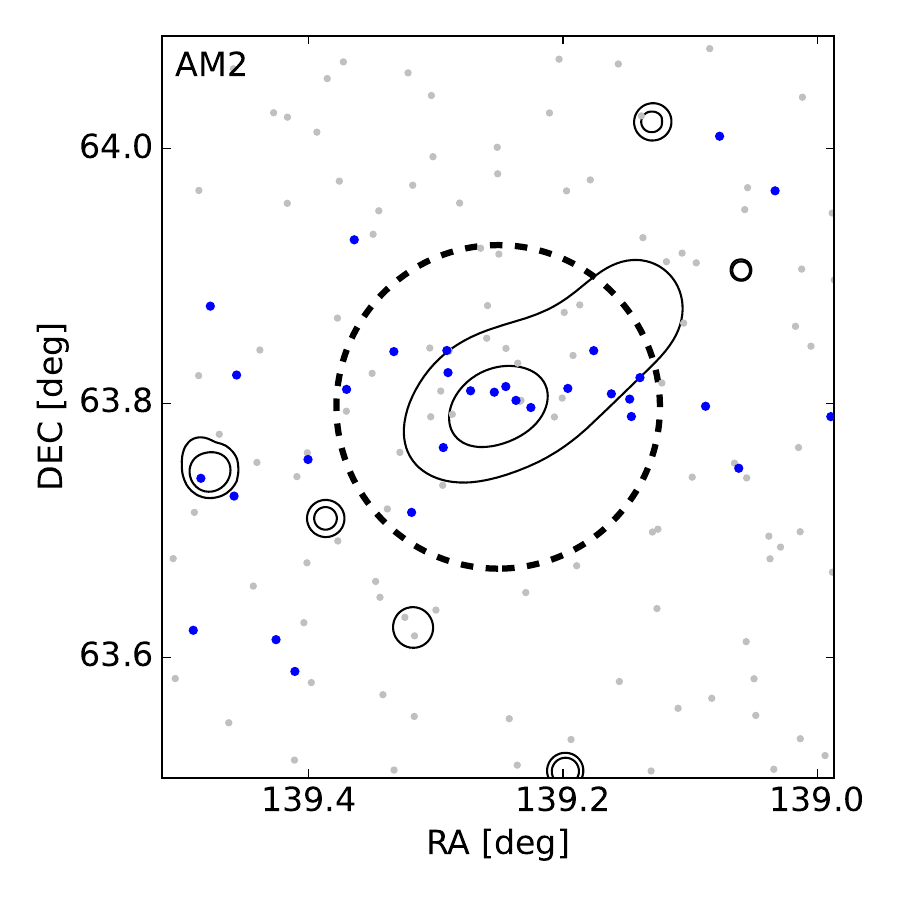}%
\includegraphics[trim=0 0 0 0,clip,width=4.5truecm]{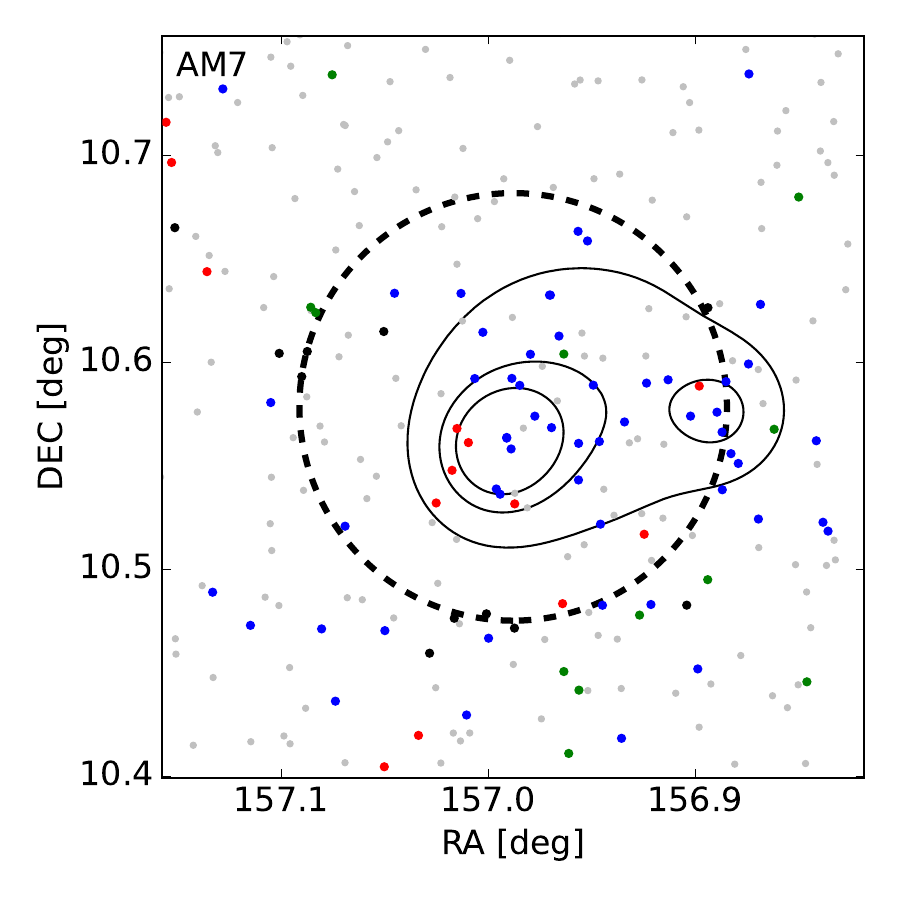}
\includegraphics[trim=0 0 0 0,clip,width=4.5truecm]{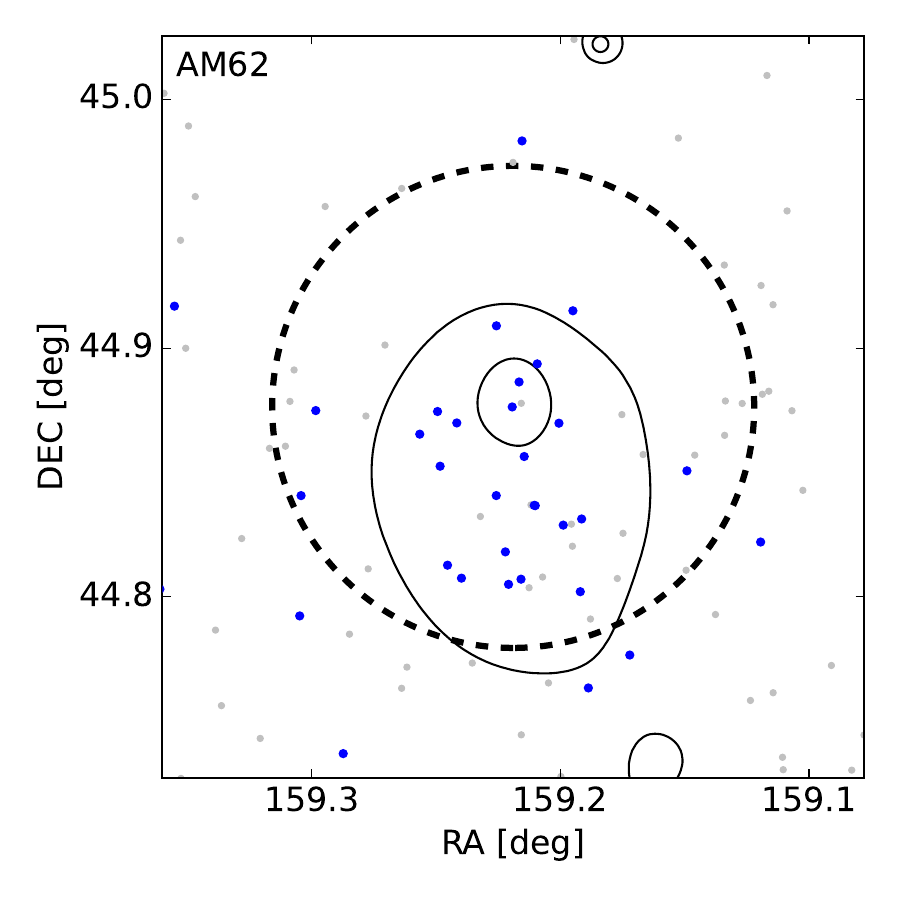}%
\includegraphics[trim=0 0 0 0,clip,width=4.5truecm]{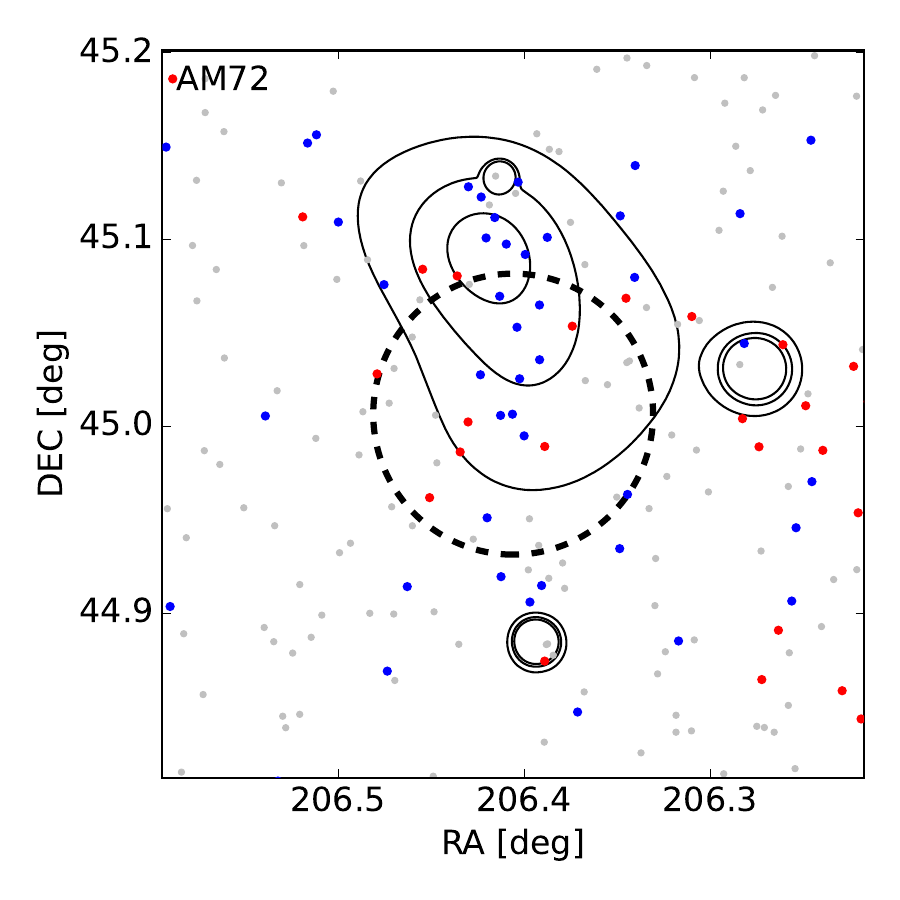}}%
\caption[h]{Spatial distribution of galaxies with spectroscopic redshifts, color-coded by membership to each redshift peak (blue points indicate the peak with the largest cardinality), with X-ray contours overlaid. No clusters are projected along the same line of sight, although in some fields (AM2 and AM72 panels) a second cluster is present within the shown field of view (undetected here in AM2 field of view because of its low cardinality). The dashed circle marks 1 Mpc at the cluster redshift.
}
\label{fig:radecz}
\end{figure*}

\begin{figure}
\centerline{\includegraphics[trim=20 0 45 0,clip,width=9truecm]{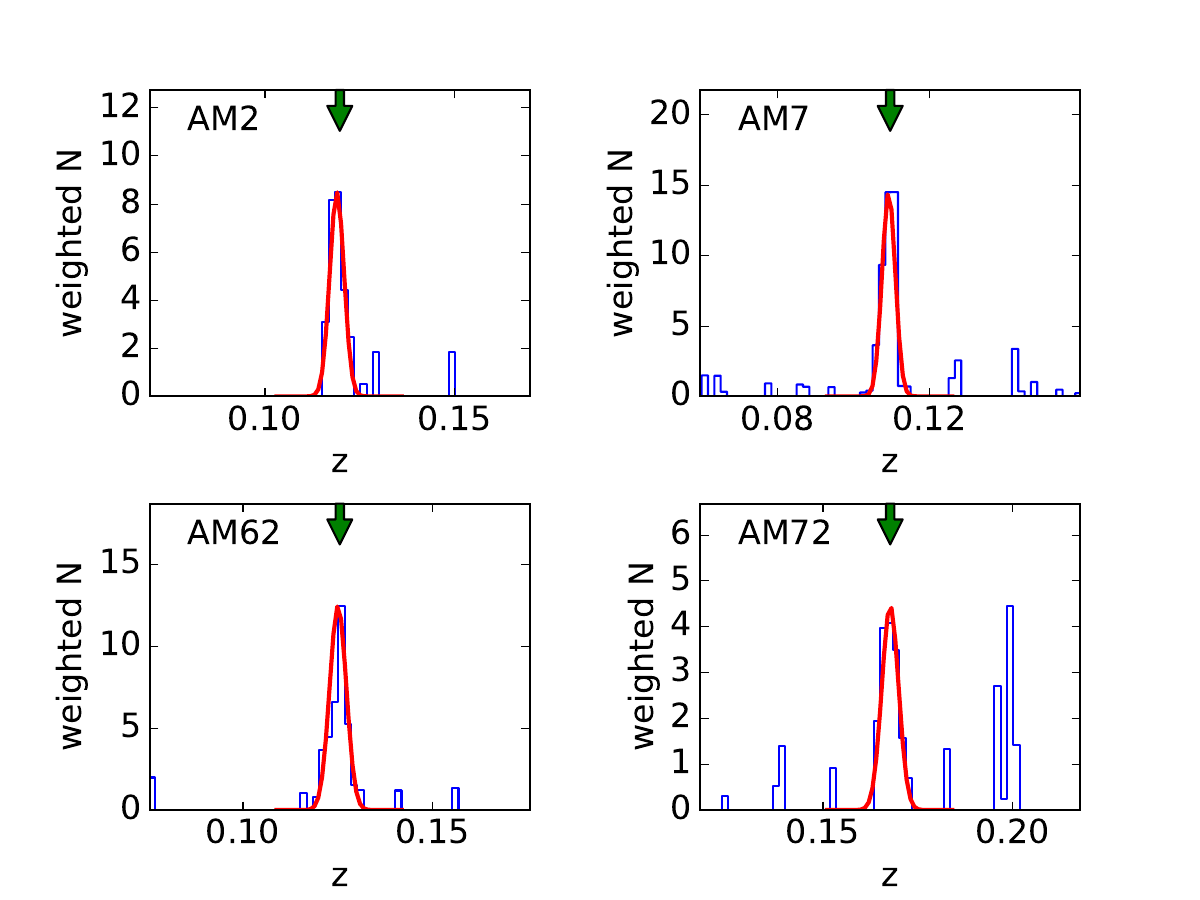}}%
\caption[h]{Weighted redshift distribution of galaxies along the lines of sight toward the four studied clusters. The arrow indicates the derived cluster redshift. The unimodal shape of the histograms shows that three of the four lines of sight are free from significant foreground or background contamination, whereas the line of sight toward AM72 is partially affected by a secondary structure at 1.3~Mpc from the cluster, with negligible impact for our analysis (see text). The narrow width of the main redshift peak ($\sim 500-700$~km/s) indicates that we are not observing multiple groups in a filament aligned along the line of sight.
}
\label{fig:histoz}
\end{figure}

\subsection{X-ray measurements}

We observed the four clusters with the \textit{Swift} X-ray Telescope (XRT), a focusing X-ray instrument with an effective area of 140~cm$^2$, a $24'$ field of view, an energy range of 0.2--10~keV (Burrows et al. 2005), and an angular resolution of $18''$ (half-power diameter) approximately constant across the field (Moretti et al. 2005). The low-Earth orbit of the mission is well suited for the detection and analysis of extended, low-surface-brightness sources such as those considered here.  In fact, for a given exposure time, XRT is approximately 1.5 times more sensitive to extended surface brightness emission than XMM-Newton, and by an even larger factor compared to Chandra (Mushotzsky et al. 2019; Walker et al. 2019).

For AM62 we obtained a deeper single pointing, co-centered with the observations from our early program. For the remaining clusters we acquired mosaic observations centered on the newly determined candidate centers. AM2 and AM7 were observed with four-point square mosaics, with pointings separated by $\sim 10'$, while AM72 was observed with a three-point triangular mosaic with side lengths of $11'$.  \textit{Swift}'s dithering around the nominal target coordinates distributes the exposure over multiple offsets, producing mosaics that broadly follow the planned pattern with minor variations. The resulting mosaics cover up to approximately $0.5 \times 0.5$~deg$^2$. 

The X-ray data were processed using the standard screening pipeline at the UK \textit{Swift} Science Data Centre, producing cleaned event files and exposure maps. The average exposure is 22.7, 19.5, 47.0, and 17.9~ks for AM2, AM7, AM72, and AM62, respectively.

To derive X-ray luminosities, we measured background-subtracted count rates within circular apertures of 300~kpc 
radius, accounting for spatial variations in exposure time, flagged regions, and standard instrumental corrections, following the procedure described in Andreon \& Moretti (2011). The commonly adopted 0.5--2.0~keV band exhibits a strongly varying background in the AM72 mosaic. In particular, more recent observations located predominantly in the northern region show enhanced background emission in the 0.5--0.8~keV band, possibly related to orbital conditions, leading to a sharp spatial gradient. For AM72 we therefore adopt the 0.8--2.4~keV band for the analysis, while the standard 0.5--2.0~keV band is used for the remaining clusters. Binned exposure-uncorrected X-ray images are shown in Figs.~\ref{fig:AM2Xray} to~\ref{fig:AM72Xray}.

The [0.5--2]~keV X-ray luminosities were derived from the measured count rates assuming an APEC plasma model (Smith et al. 2001) with temperature $T = 5$~keV, metallicity 0.3 solar, and total Galactic hydrogen column density $n_{\rm H}$ from Willingale et al. (2013). Apertures are centered on the X-ray peak. Alternative centering choices would yield lower luminosities, further reinforcing the underluminous nature and low--surface--brightness of these systems.

For visualization purposes, X-ray contours were derived from exposure-corrected event files in the adopted energy bands using adaptive smoothing with \texttt{CIAO csmooth} (Ebeling et al. 2006), retaining only features with $S/N > 3$, following a suggestion by C. Sarazin. Because \texttt{csmooth} assumes integer-valued inputs and does not account for spatially varying exposure, we adopted the following approximation: count-rate images were multiplied by 100 times the mean exposure, rounded to the nearest integer, and masked in regions with exposure below 30\% of the mean. The $S/N$ threshold was scaled by the square root of the same factor. This procedure enforces an approximately constant $S/N$ in count rate rather than in counts. The approximation is adequate for display purposes and negligible compared to the intrinsic faintness of the extended emission and point-source contamination, particularly in AM72 (Fig.~\ref{fig:AM72Xray}). The resulting adaptively smoothed contours are overlaid on binned exposure-uncorrected images in Figs.~\ref{fig:AM2Xray} to~\ref{fig:AM72Xray} and Fig.~\ref{fig:maps}. Individual clusters are discussed after presenting the full multiwavelength dataset.

As noted above, we use data processed with the standard UK \textit{Swift} Science Data Centre pipeline rather than the custom pipeline adopted in our previous works (e.g. Andreon et al. 2016, 2019, 2022), which does not support mosaic observations. A re-analysis of the single-pointing AM62 data with our custom pipeline yields consistent X-ray luminosities.

\begin{figure*}
\centerline{\includegraphics[trim=20 0 40 30,clip,width=4.5truecm]{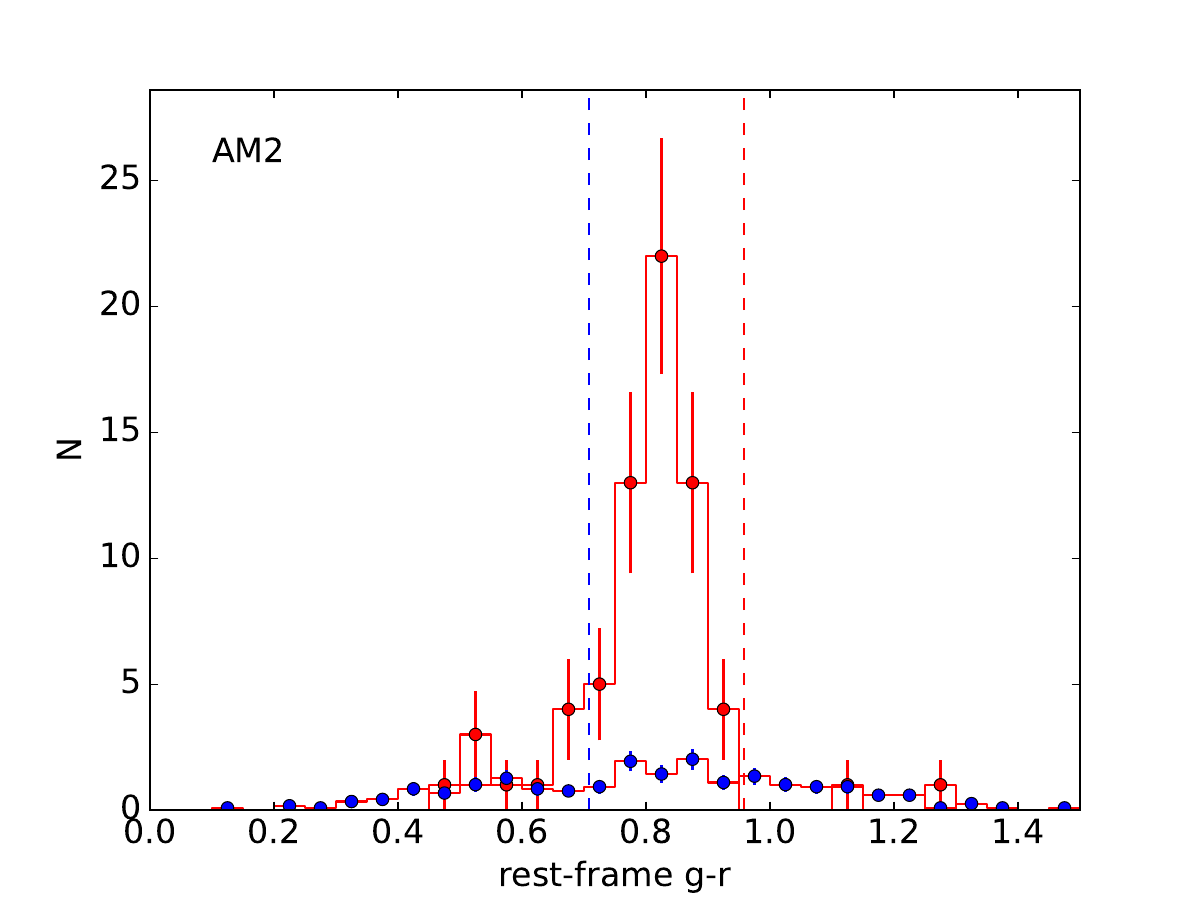}\includegraphics[trim=20 0 40 30,clip,width=4.5truecm]{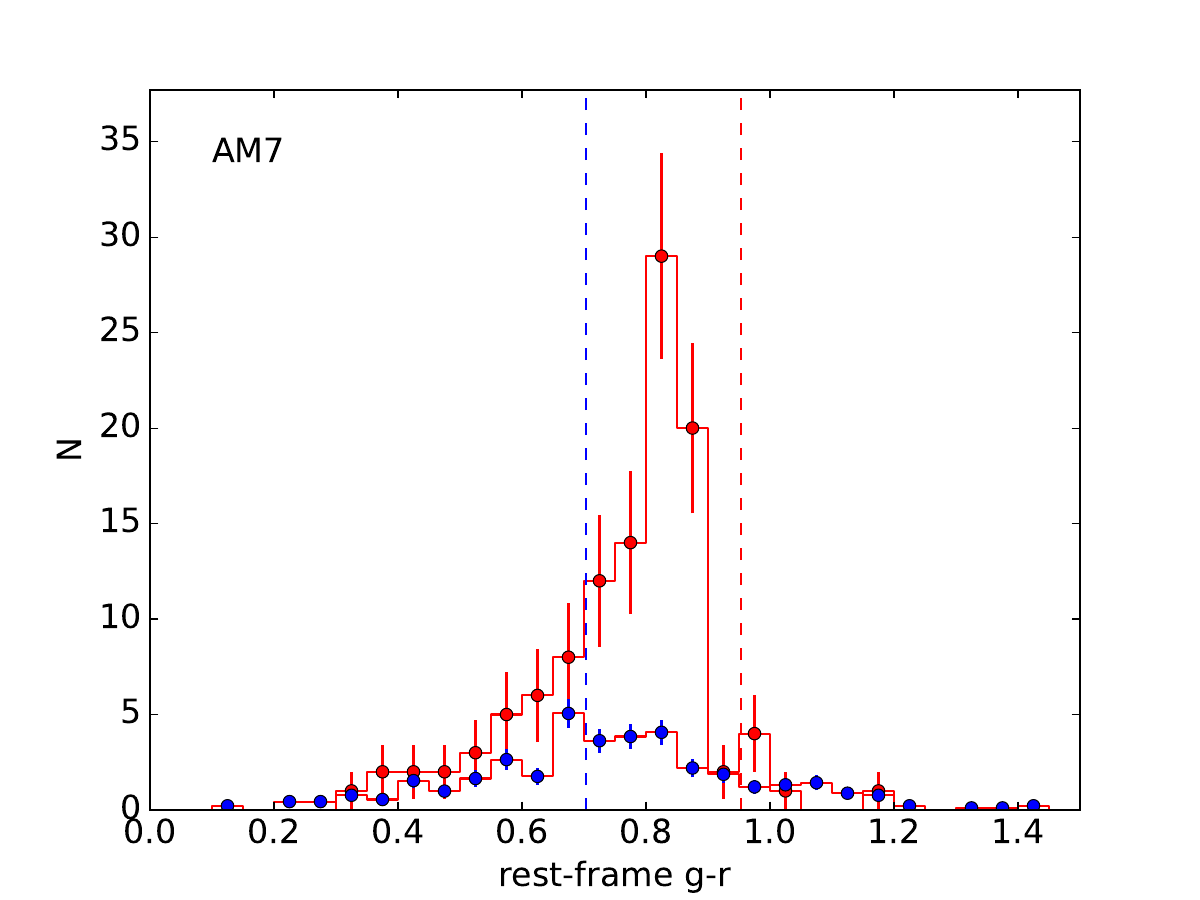}
\includegraphics[trim=20 0 40 30,clip,width=4.5truecm]{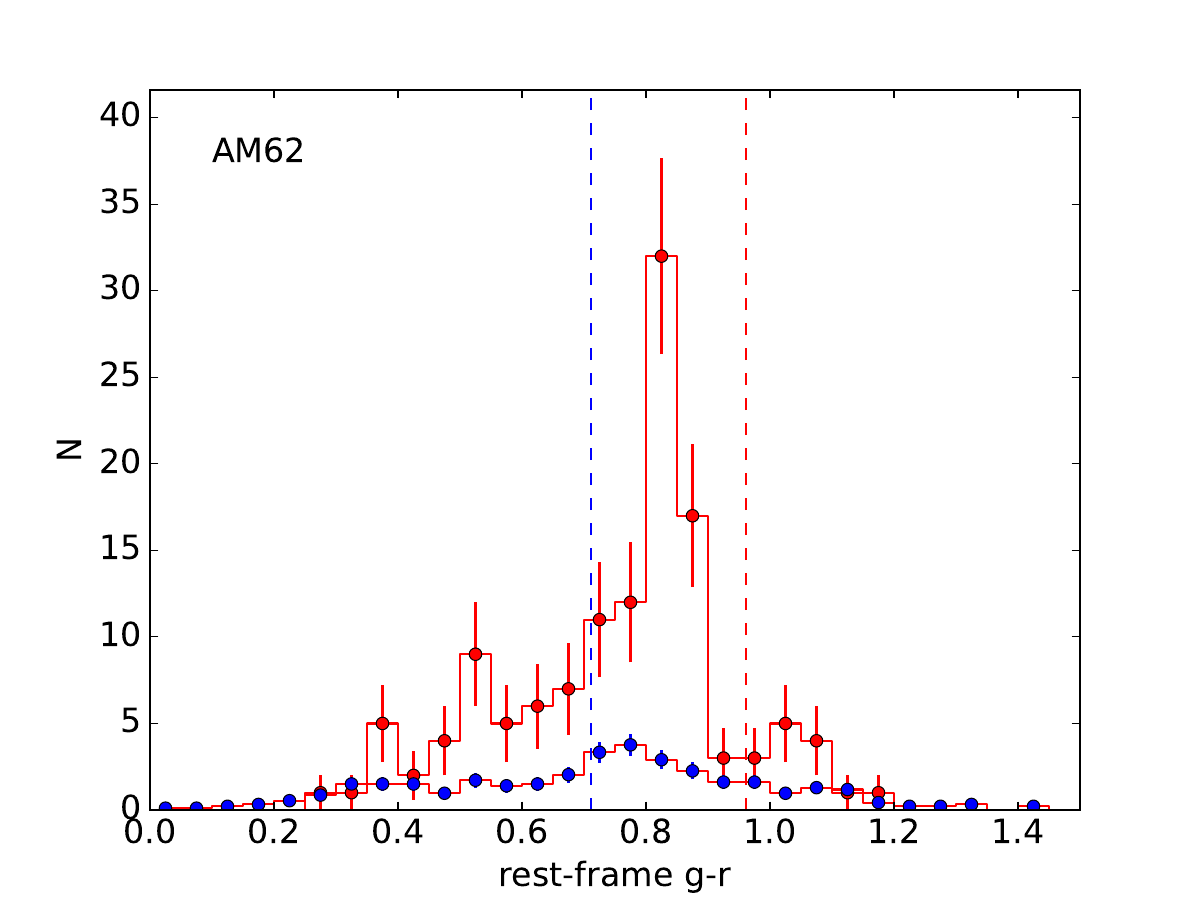}\includegraphics[trim=20 0 40 30,clip,width=4.5truecm]{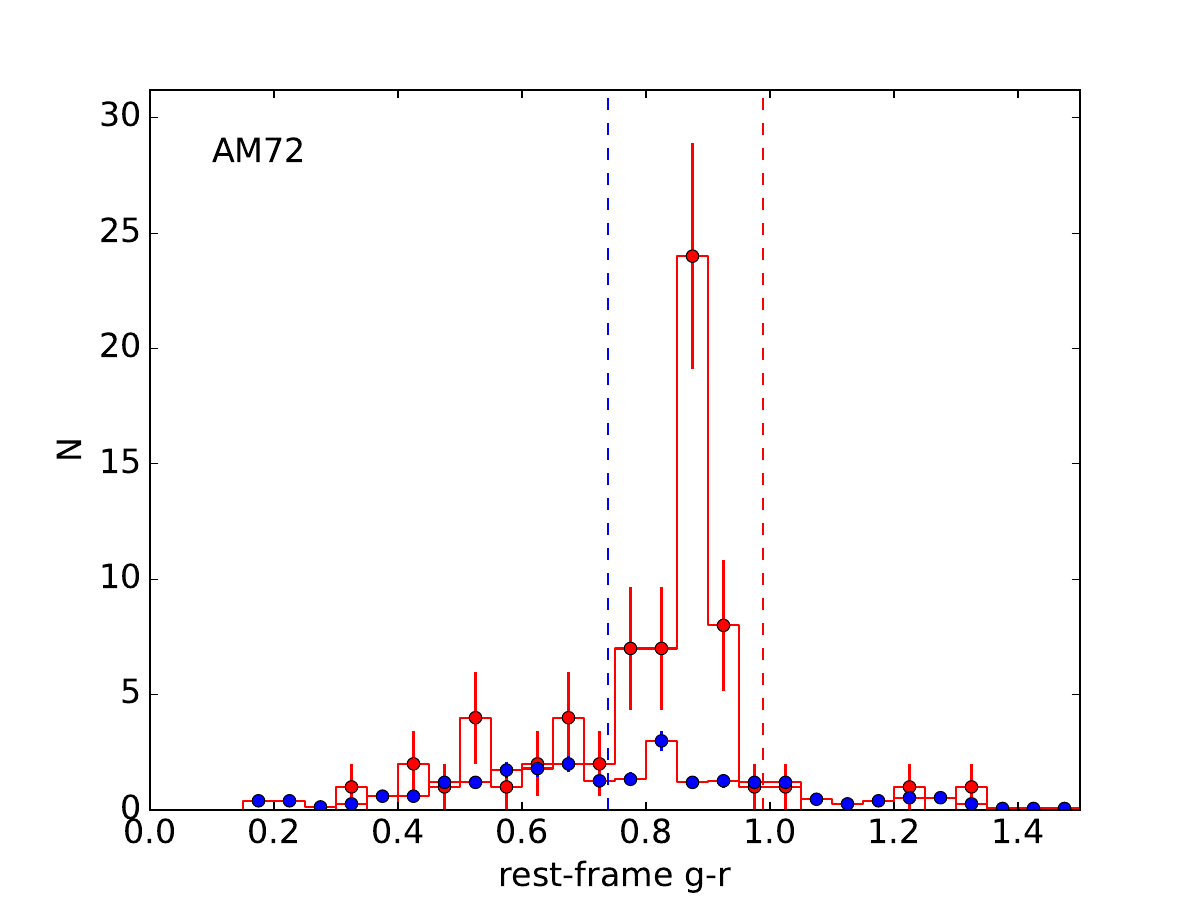}}%
\caption[h]{Color distribution of galaxies within $r_{200,\rm gal}$ (red histogram) compared to a control area (normalized to the cluster solid angle, blue histogram) for the four clusters. The vertical lines indicate the expected color range of the red sequence. Only galaxies brighter than $M^e_V=-20$~mag, assuming all galaxies are at the cluster redshift, are included. The unimodal distributions indicate that there are no additional clusters along the line of sight separated by $\Delta z > 0.02$, the resolution limit of this method.
}
\label{fig:coldistr}
\end{figure*}

\begin{table*}
\caption{Cluster sample and summary of analysis results.}
\begin{tabular}{lrrrrrrrrrrrrrr}
\hline\hline
ID & \multicolumn{1}{c}{R.A.} & Dec &  z$_{\mathrm spec}$ &   $n_{200,A16}$ & err & $\log L_{\mathrm X}(<500\mathrm{kpc})$ & err & $\log \mu_{300}$ & err\\
 & \multicolumn{2}{c}{J2000} & & &  & erg s$^{-1}$ & dex & erg s$^{-1}$ Mpc$^{-2}$ & dex\\
\hline
 (1) & (2) & (3) &  (4) & (5) & (6) & (7) & (8) & (9) & (10) \\
\hline
AM2 & 139.251 &  63.797  &  0.1197  & 48 & 8 & 43.23  & 0.02  & 43.42  &  0.04         \\
AM7 & 156.988 &  10.5786 &  0.1097  & 66 & 9 & 43.11  & 0.03  & 43.48  &  0.03       \\
AM62& 159.219 &  44.8764 &  0.1255  & 64 & 9 & 43.15  & 0.03  & 43.50  &  0.04      \\
AM72& 206.406 &  45.0066 &  0.1677  & 38 & 7 & 42.80   & 0.05  & 43.11  &  0.05     \\
\hline
\end{tabular}
\hfill \break 
{\bf Notes.} The table provides the following information: cluster ID (Col. 1); coordinates of the X-ray peak (RA and Dec, Cols. 2 and 3); spectroscopic redshift z$_{\mathrm spec}$ (Col. 4);  richness $n_{200,A16}$ within the radius inferred from galaxy counts, along with its uncertainty (Cols. 5 and 6), (log of) X-ray luminosity within 500 kpc with its uncertainty (Cols. 7 and 8) and (log of central) X-ray surface brightness (within 300 kpc) with its uncertainty (Cols. 9 and 10).
 \hfill
\label{tab1}
\end{table*}

\subsection{Redshift measurements and redshift histograms}
\label{sec:z}

We use spectroscopy from DESI (DESI Collaboration 2026), SDSS (Almeida et al. 2023), and literature redshifts compiled in NED. Duplicate measurements were removed. Peaks in the redshift distribution were searched within a $\pm 1500$~km~s$^{-1}$ window placed at the redshift of every galaxy. The search radii were set to (7, 10, 6, 8)$'$ for AM2, AM7, AM62, and AM72, respectively, corresponding to approximately 1.5 times the X-ray extent of each cluster. Redshift peaks containing at least four galaxies were retained. Peaks were merged using a $\pm 2000$~km~s$^{-1}$ linking length, retaining the peak with the larger number of members.

Figure~\ref{fig:radecz} shows the spatial distribution of galaxies with spectroscopic redshifts, color-coded by membership to each redshift peak (blue marks the peak with the largest number of members), overlaid with X-ray contours. No cases of two clusters projected along the same line of sight are identified, although additional clusters are present elsewhere in the surveyed field. We adopt as cluster redshift the redshift of the galaxy corresponding to the peak with the largest membership; these values are listed in Table~1. A simpler selection based on the peak within 1~Mpc of the X-ray center yields identical results.

To further test for multiple structures along the line of sight, we constructed redshift histograms toward each cluster. To mitigate the effects of sparse central sampling due to fiber collisions and the increasing projected area at larger radii (which enhances contamination), we assign each galaxy a radial weight that decreases linearly from 2 at the X-ray center to 0 at 1~Mpc, corresponding to an average weight of 1. This weighting scheme down-weights galaxies at large projected distances and effectively restricts the analysis to the X-ray emitting region. Weighted galaxy counts per redshift bin are then computed. The resulting histograms are shown in Fig.~\ref{fig:histoz}. Three clusters exhibit a single dominant peak. AM72 shows an additional secondary peak corresponding to a background cluster located at a projected distance of $\sim 1$~Mpc, also visible in the X-ray image (at RA = 206.25; Fig.~\ref{fig:AM72Xray}) and discussed in Sect.~\ref{sec:AM72}.

The redshift histograms were fitted with a Gaussian profile (red curves in Fig.~\ref{fig:histoz}). The measured velocity spreads range from 520 to 660~km~s$^{-1}$, consistent with individual clusters rather than superpositions of multiple structures at different redshifts. Apart from the spatially distinct secondary peak in AM72, no additional peaks separated by more than 500~km~s$^{-1}$ are detected, as would be expected for a filament composed of multiple groups projected along the line of sight. Other redshift intervals not shown also reveal no significant overdensities.

We note that spectroscopic completeness decreases with redshift, so additional systems at substantially different redshifts could remain undetected in spectroscopy alone. Such systems are more robustly identified through the color-based richness analysis presented in Sect.~\ref{sec:richness}, which includes galaxies irrespective of spectroscopic coverage and is therefore insensitive to spectroscopic incompleteness.

In summary, the redshift analysis excludes the possibility that the observed systems are projections of two or more clusters along the same line of sight, or segments of a filament seen in projection within $\Delta z \sim 0.1$. The only additional structure identified is the background cluster located $\sim 1$~Mpc from AM72.

\subsection{Richness measurements}
\label{sec:richness}
Cluster richnesses are derived from SDSS photometry (Almeida et al. 2023), following the methodology of Andreon \& Radovich (2026), which builds on Andreon (2015, 2016) with minor updates and constitutes one of the two richness estimators adopted by the Euclid Collaboration (Euclid Collaboration: Mellier et al. 2025). In brief, we count galaxies on the red sequence that are brighter than the passively evolved threshold $M^e_V = -20$~mag within a radius $r_{200}$.

The radius $r_{200}$ is determined iteratively from the richness measured within $r_{200}$ itself, following previous work and in analogy with the standard procedure used for the $Y_X$ observable. Our operational definition of "on the red sequence" includes galaxies within 0.1~mag redward and 0.15~mag blueward of the fitted color--magnitude relation. As in Andreon (2015, 2016), we adopt \texttt{ModelMag} and \texttt{cModelMag} magnitudes for colors and total fluxes, respectively. The background contribution is estimated in an annulus between 3 and 7~Mpc, with corrections for contamination by unrelated structures. To reduce bias from localized overdensities or underdensities, the annulus is divided into octants and the two octants with the highest and the two with the lowest galaxy counts are excluded from the background estimate.

As in Andreon \& Radovich (2026) and in the Euclid analysis (Euclid Collaboration: Mellier et al. 2025), we adopt a stricter blueward color cut of 0.15~mag, compared to the 0.2~mag used in our earlier works, to reduce contamination from foreground systems along the line of sight. As shown in Andreon \& Radovich (2026), an even more restrictive 0.1~mag blueward selection yields consistent redshift estimates for clusters not affected by foreground contamination.

To test for additional clusters projected along nearby lines of sight that may have escaped detection in Sect.~\ref{sec:z} because of  spectroscopic incompleteness, we computed the rest-frame $g-r$ color distribution within $r_{200}$ for galaxies brighter than $M^e_V = -20$~mag, assuming all galaxies lie at the target cluster redshift (Fig.~\ref{fig:coldistr}). Clusters separated by $\Delta z > 0.02$ would produce bimodal color distributions (Gladders \& Yee 2000, Andreon \& Berge 2012). The unimodal distributions observed for all four systems indicate negligible contamination from other massive structures and support the robustness of the richness measurements.

\begin{figure}
\centerline{\includegraphics[width=8truecm]{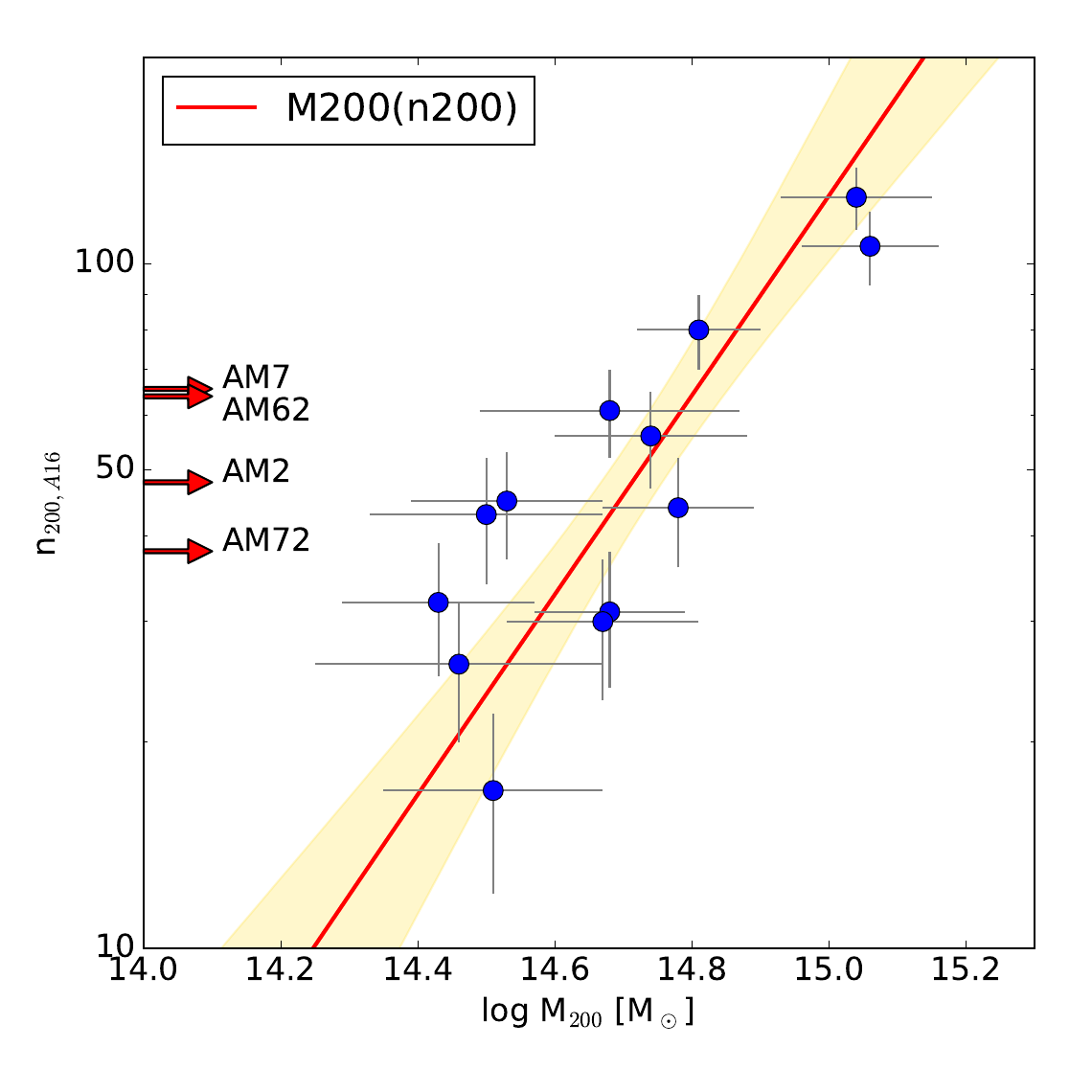}}
\caption[h]{
Richness--mass scaling relation. Blue points and the solid line with yellow shading show the mean richness--mass model with 68\% confidence region from the Andreon \& Radovich (2026) sample. The four clusters studied in this work are indicated by arrows; their richness corresponds to $\log M/M_\odot = 14.6 \pm 0.1$.}
\label{fig:Mrich}
\end{figure}

Figure~\ref{fig:Mrich} shows the richness--mass scaling relation derived for the weak-lensing-selected sample of Andreon \& Radovich (2025b), with the four clusters analyzed here indicated by arrows. As quantified in that study and evident from the figure, $n_{200,\mathrm{gal}}$ is a low-scatter mass proxy, with an intrinsic scatter of 0.11~dex, improving upon the earlier estimate of Andreon (2016). Using this calibration, the four clusters studied here have masses of $\log M/M_\odot = 14.5$ to  $14.8$.

For comparison, the Andreon \& Radovich (2026) sample (solid points) includes all the most massive clusters, defined as those with the largest weak-lensing shear signal, within a 150~deg$^2$ region and in the redshift range $0.14 < z < 0.42$. The clusters analyzed here lie close to the mean richness of that sample, confirming that they belong to the high-mass population.

\subsection{Planck undetection}

Despite their low redshift and high richness, none of the four clusters is detected by Planck (Planck Collaboration 2016b). We verified that all systems lie at least 1.5$^\circ$ from masked regions (Planck Collaboration et al. 2016a), ruling out masking as the cause of the non-detections.

Richness-based mass estimates place these clusters above, but near the lower end of, the nominal Planck detection threshold. Therefore, the lack of detection can plausibly be attributed to intrinsic scatter between richness and Compton $Y$. Inspection of the improved Compton-$Y$ maps from Chandran et al. (2023) reveals tentative signal at the location of AM2, and possibly AM62 as well, although for the latter the positional agreement is poor.

\subsection{Discussion about individual objects}

\subsubsection{AM2}

AM2 is at $z=0.120$ (Fig.~\ref{fig:histoz}). In the field, there is another cluster at $z=0.139$, located 1.2~Mpc (projected) NW of AM2, where the X-ray contours become more elongated (Figs.~\ref{fig:AM2Xray}, \ref{fig:maps}, and \ref{fig:radecz}). This secondary cluster is weak, lies outside our apertures of interest, and therefore its impact on the AM2 measurements is negligible. It has only three spectroscopically concordant redshifts, below the threshold used for the spectroscopic analysis in Sec.~2.3, and is thus not detected there. Apart from this, no other significant groups are present along the line of sight, and the redshift histogram width is inconsistent with a filamentary structure (Fig.~\ref{fig:histoz}).
The low signal-to-noise ratio of the X-ray data used in Andreon \& Moretti (2011), combined with the lack of spectroscopic information available at the time, prevented the identification of AM2 as an additional case of a miscentered maxBCG cluster. The currently adopted center is offset by 3' (about 0.4 Mpc) from the position reported in the maxBCG catalog.

The X-ray emission of AM2 is clearly elongated along the NW--SE direction (Fig.~\ref{fig:AM2Xray}), with an axis ratio of 1.5, measured at the second shown isophote. The apparently abrupt discontinuity in surface brightness toward the NW corresponds to a sharp change in exposure time in the mosaic and is therefore not a shock feature.

AM2 corresponds to Abell 764, also known as ZwCl 0913.1+6358, and was previously serendipitously detected in X-rays by David, Forman \& Jones (1999) as an off-axis source in a ROSAT PSPC pointing of NGC 2805.

\subsubsection{AM7}

AM7, at $z=0.110$, is a bimodal cluster (see Fig.~\ref{fig:AM7Xray}) with approximately 30 spectroscopic members within the faintest plotted X-ray isophote, and it exhibits two X-ray clumps at the same redshift (Fig.~\ref{fig:radecz}). The western X-ray clump is 0.2~dex fainter than the eastern one within a 300~kpc radius aperture.  

Approximately 2$'$ southeast of AM7, there is a group of five galaxies at $z=0.321$ (Fig.~\ref{fig:radecz}), with no obvious X-ray excess at their location. Apart from this, no other significant groups are present along the line of sight, and the redshift histogram width is inconsistent with a filamentary structure.  

AM7, also known as ZwCl 1025.2+1050, is the only cluster in our sample that falls within the eROSITA Western footprint (Merloni et al. 2024). Despite its large richness and proximity, only 20 net photons are detected within 3$'$ of the cluster position in the 0.5--2.4~keV band in the Data Release 1 images. Consequently, it is unsurprising that AM7 is absent from the eROSITA cluster catalog (Bulbul et al. 2024).

\subsubsection{AM62}

AM62 is a morphologically disturbed cluster, with strongly asymmetric X-ray emission along the North--South direction (Fig.~\ref{fig:AM62Xray}). The southern region is about 0.3~dex fainter than the peak within a 300~kpc radius aperture, spatially
coincident with a galaxy overdensity (Fig.~\ref{fig:maps}) and also closely traced by the cluster members Fig.~\ref{fig:radecz}.

No other clusters are present along the line of sight (Figs.~\ref{fig:radecz}, \ref{fig:histoz}, and \ref{fig:coldistr}). The tight velocity distribution around the cluster redshift (Fig.~\ref{fig:radecz}) confirms that AM62 is a single system rather than multiple structures projected along nearby lines of sight.

\subsubsection{AM72}
\label{sec:AM72}

The AM72 field of view contains a second cluster: approximately 8~arcmin (1.3~Mpc) west of AM72 lies MaxBCG J206.24877+45.01110 at $z = 0.199$, clearly visible both in the X-ray map (Fig.~\ref{fig:AM72Xray}) and in the redshift histogram (Fig.~\ref{fig:histoz}). 
Its much larger distance compared to the X-ray apertures used for AM72 makes its impact negligible for our X-ray aperture photometry. Its richness, derived following the same procedure as for the target clusters, indicates that it is about 2.5 times poorer than AM72. Its presence does not bias the AM72 richness measurement: its lower richness, higher redshift, and the resulting faintness, and color difference of most of its galaxies prevent them from being counted as AM72 members.

AM72 itself, at $z = 0.122$, shows a structure elongated in the North--South direction as traced by galaxies (Fig.~\ref{fig:maps}), spectroscopic members (Fig.~\ref{fig:radecz}), and X-ray emission (Fig.~\ref{fig:AM72Xray}). It is composed of at least two clumps, both confirmed spectroscopically at the same redshift, separated by roughly 1~Mpc. The southern clump is the X-ray brightest, bending the outermost X-ray contours toward the south. The northern X-ray emission is instead a blend of extended emission and point sources, one of which (the closest to the center of the inner isophote) is a confirmed cluster member. The overall X-ray brightness of the system is extremely low (Sec.~2.2).

\begin{figure}
\centerline{\includegraphics[width=8truecm]{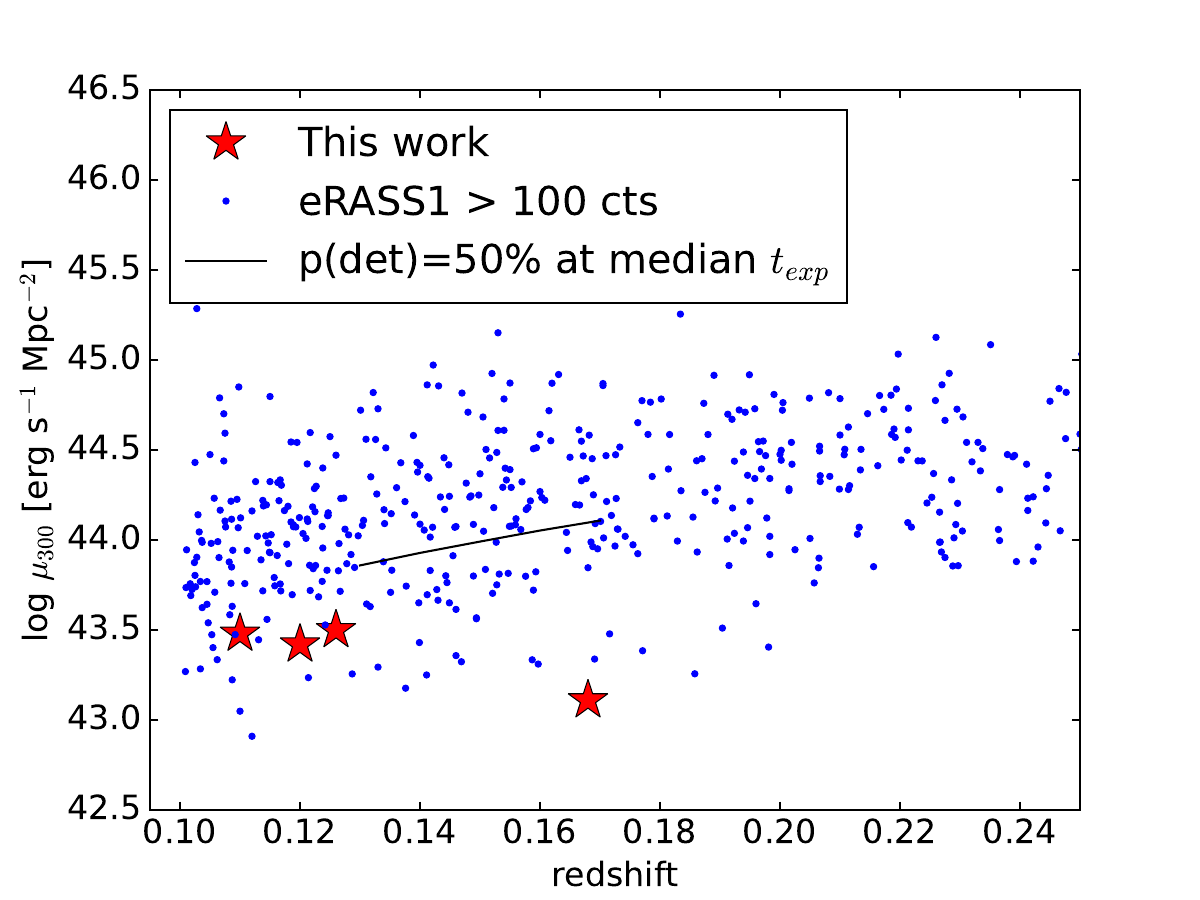}}
\caption[h]{
Distribution in the brightness--redshift plane of eROSITA clusters (dots) and our targets (stars). The curve shows the 50\% eROSITA detection probability (Ghirardini et al. 2025). Our targets fall below the 50\% detection probability and are at the faintest end or below it, at fixed redshift of the eROSITA  distribution indicating that they are uncommon in that sample.}
\label{fig:mu_z}
\end{figure}

\subsection{Collective analysis}
\label{sec:collective}

In the previous sections, we have shown that our sample consists of bona fide rich, and therefore massive, clusters with strongly disturbed morphologies in both the optical and X-ray domains.

Figure~\ref{fig:mu_z} shows the X-ray brightness within a 300~kpc radius for the studied clusters (stars), compared to clusters in the cosmological sample of the eROSITA Data Release 1 (Bulbul et al. 2024; Ghirardini et al. 2024). For the latter, we consider only clusters in the cosmological sample with at least 100 counts within the 300~kpc aperture, for which $\log \mu_{300}$ is typically measured with $\sim 10\%$ uncertainties. As a comparison, our clusters have typically $\sim 450$ photons within the same aperture. At fixed redshift, the clusters in our sample lie at the low end of the eROSITA cluster distribution or below it. The lower boundary of the eROSITA distribution is strongly dependent on the adopted minimum count threshold: when increasing the threshold to 200 counts, all eROSITA clusters fainter than our brightest target are removed.
Our systems are therefore poorly represented in the eROSITA cosmological sample, mainly because the 50\% detection probability at the median survey depth corresponds to brighter systems (solid line in Fig.~\ref{fig:mu_z}; Ghirardini et al. 2024).
 
\begin{figure}
\centerline{\includegraphics[trim=0 0 0 0,clip,width=9truecm]{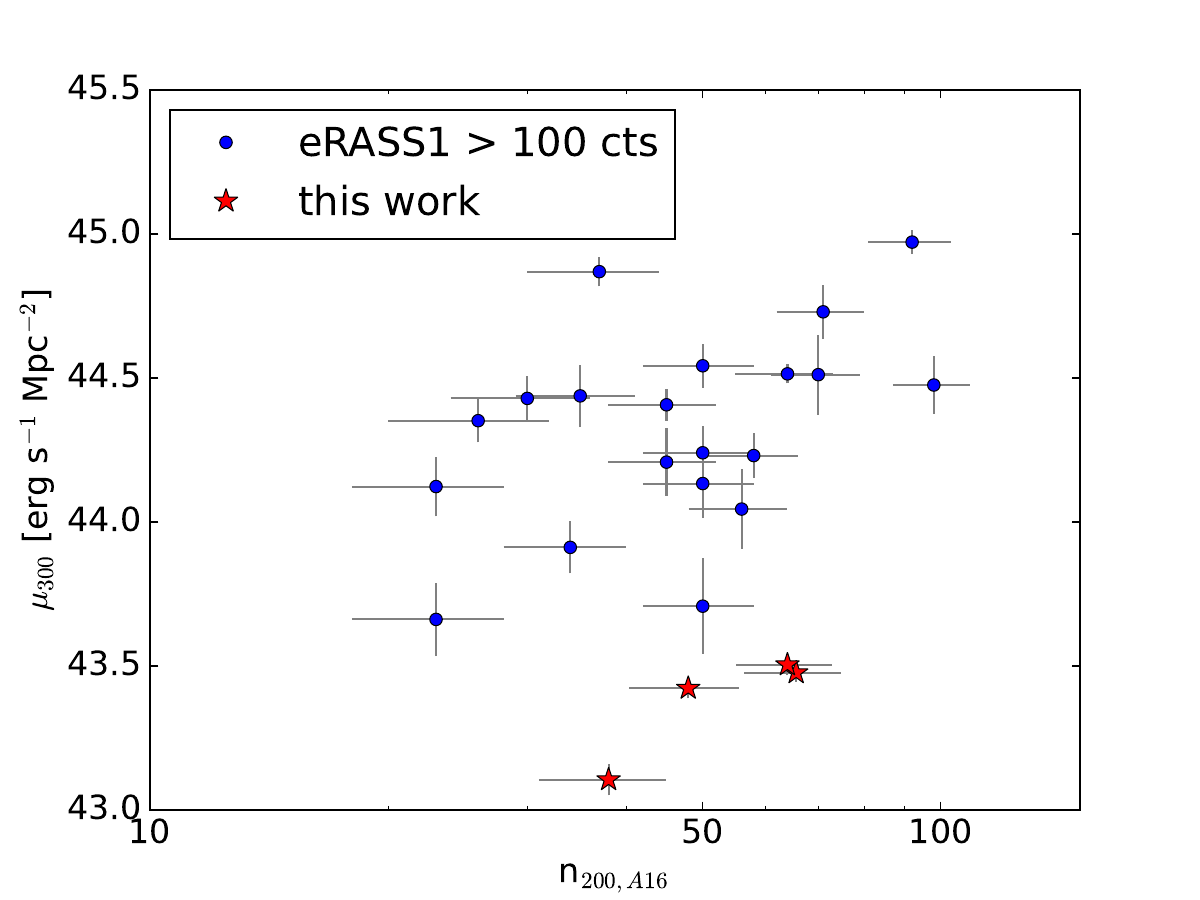}}
\caption[h]{Richness versus X-ray brigthness within 300 kpc for our sample (stars) and for an ICM-selected comparison sample (see text for details). The clusters studied here are systematically fainter by approximately one dex compared to ICM-selected systems at the same richness.
}
\label{fig:mu300_R}
\end{figure}

The above comparison presents the limit that eROSITA clusters with low $\mu_{300}$ may correspond either to massive, low--surface--brightness systems similar to those in our sample, or to lower-mass groups with more typical luminosities. Disentangling these cases requires independent mass or richness estimates, which are shown in Figure~\ref{fig:mu300_R} for clusters in common between Andreon (2016) and eROSITA systems with at least 100 counts. To increase the comparison sample, we also include clusters with declination $\delta > 32^\circ$ avoided by the cosmological sample. While we acknowledge that the resulting cross-matched sample lacks a formally defined selection function, we emphasize that the Andreon (2016) sample was constructed using solely a richness threshold (richness $> 21.6$). The selection process was entirely blind to X-ray data, meaning no clusters were excluded on the basis of low X-ray surface brightness as no X-ray data was utilized to exclude objects.
At fixed richness, our clusters are systematically fainter in X-rays by approximately one dex compared to ICM-selected systems,  suggesting that most of the eROSITA low--surface--brightness systems are groups. The wide range of X-ray brightness at fixed richness revealed by our sample would be difficult to infer using only eROSITA clusters, as the latter exclude faint systems like those studied here.  Although estimating the fraction of clusters poorly represented in the eROSITA survey is beyond the scope of this work, given the limited sample size, the missed population is likely small but not negligible. Four clusters out of 38 in the parent sample, possibly supplemented by three additional clusters with similar central luminosities that were not observed and are therefore likely at the very limit, or below, of the detectability in current X-ray surveys, already correspond to $\sim$20\% of the sample. To this, one should add the possibility that some fraction of the remaining clusters may also fall below current detection thresholds.

We note that, in order for these systems to be consistent with a group-like X-ray brightness of $10^{43}$~erg/s/Mpc$^2$, their richness would need to be as low as $\sim 10$, implying a contamination level roughly four times larger than the cluster population itself. This scenario is already ruled out by the spectroscopic and color data.
As an additional test, we computed a conservative lower limit to the cluster richness by counting only spectroscopically confirmed red-sequence galaxies within $r_{200}$ and brighter than the passively evolved threshold $M^e_V = -20$~mag. This estimate is strictly a lower bound, as it excludes galaxies that satisfy the color, luminosity, and radial criteria but lack spectroscopic information. Even under this conservative approach, each target has at least 17 confirmed members. Using these lower-limit richness values, the clusters remain X--ray--faint relative to the richness--X-ray brightness relation (Fig.~\ref{fig:mu300_R}), demonstrating that their X-ray underluminosity cannot be attributed to an anomalously high richness.

\begin{figure}
\centerline{\includegraphics[trim=20 0 40 0,clip,width=7truecm]{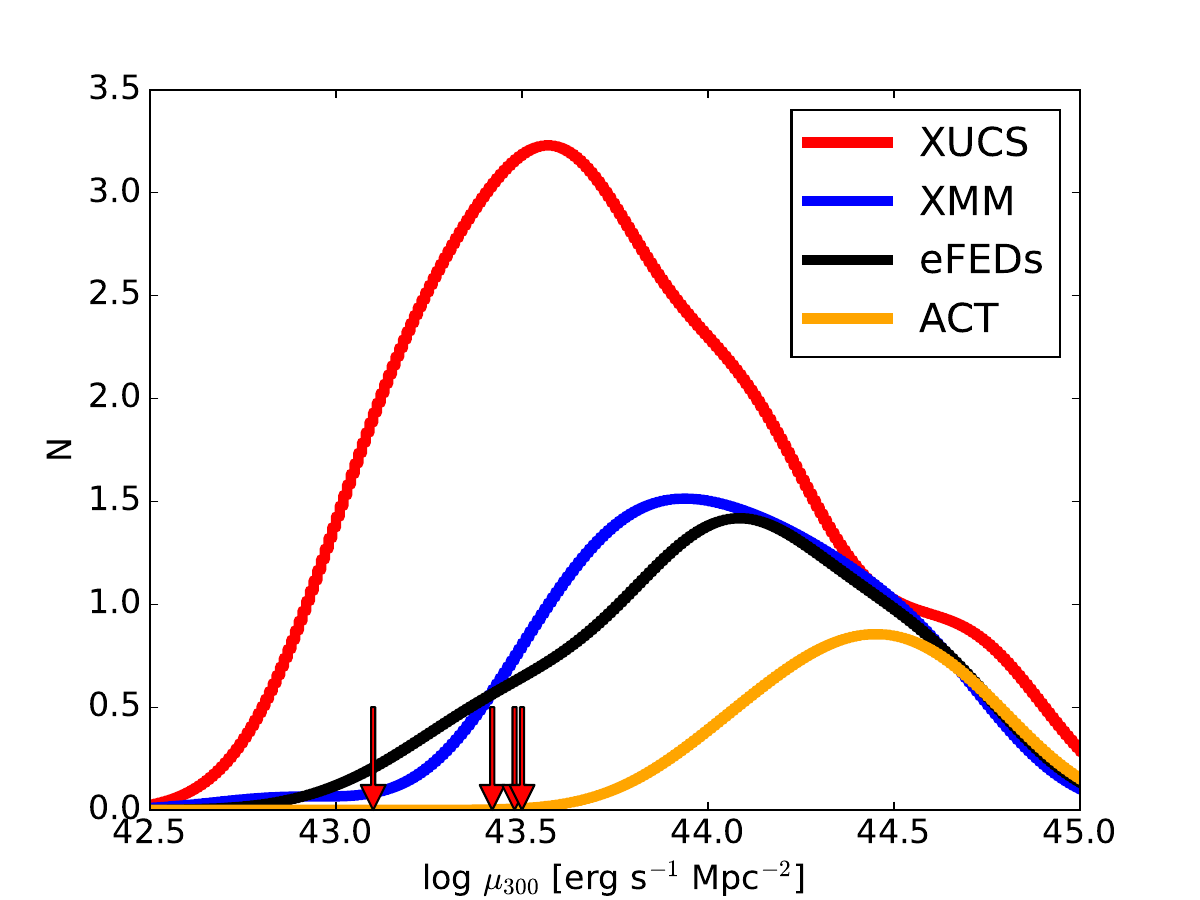}}%
\caption[h]{X-ray brightness within a 300~kpc radius aperture for our sample (arrows) and for comparison clusters from Andreon et al. (2024). Compared to ICM-selected samples (XMM, eFEDS, ACT in the figure), the studied clusters rank among the lowest in surface brightness, and are instead typical of the X-ray Unbiased Cluster Sample (XUCS, shown in the figure).
}
\label{fig:mudist}
\end{figure}

Figure~\ref{fig:mudist} shows the X-ray brightness within a 300~kpc radius for the studied clusters (arrows), compared to several reference samples from Andreon et al. (2024): the velocity-dispersion-selected X-ray Unbiased Cluster Sample (XUCS, Andreon et al. 2016) and three intracluster medium-selected samples (XMM-XXL: Pierre et al. 2016; eFEDS: Brunner et al. 2022; and ACT: Hilton et al. 2021). The studied clusters have brightness values close to the average of the XUCS sample and rank among the faintest systems in eFEDS, which has roughly ten times the median exposure of the eROSITA Data Release 1 survey. 
The studied clusters also rank among the faintest systems in XMM-XXL, which has a typical
exposure time of 10 ks, again indicating that the X-ray variety captured by 10 ks XMM surveys underestimates the true diversity, even for rich clusters like those in our sample.

\section{Conclusions}

We have studied four nearby, optically rich clusters that appear faint in archival X-ray data, mostly belonging to our own programs, relative to their richness, combining new X-ray observations with a field of view up to half degree in size
with archival spectroscopic and optical data to characterize their physical properties and assess potential selection biases.

All four systems are bona fide rich clusters, confirmed via spectroscopy, red-sequence photometry, and X-ray emission. They are not projections of multiple clusters along the line of sight, nor filamentary structures, as indicated by velocity spreads in the range 520--660~km/s, consistent with single clusters rather than multiple structures at different redshifts. They are also not heavily contaminated by foreground or background groups. Of course, other clusters are present in the fields, with the two closest lying at a projected distance of $\sim$1~Mpc from our targets.

The targets exhibit strongly disturbed morphologies in both the optical and X-ray domains, with elongated X-ray emission or multiple sub-clumps, highlighting dynamical activity and ongoing merging processes. Spectroscopy confirms that these various components are part of single dynamical systems.

Despite their high richness and low redshift, none of the clusters is detected in Planck or the eROSITA Data Release 1, and their X-ray brightness is at or below the  50\% detection limit.
Their X-ray luminosities are systematically lower than those of ICM-selected clusters at comparable richness, often by roughly one dex. This demonstrates that X-ray-selected samples alone may miss a $>20\%$ fraction of massive clusters with low--surface--brightness, including those that are optically rich, with inferred masses $\log M_{200}/M_\odot \sim 14.6$. The observed low X-ray luminosity of our targets cannot be explained by the targets being multiple structures seen in projection or their 
richness determination being erroneous.

These results underscore the importance of multi-wavelength approaches to cluster selection. Optical searches identify clusters that would otherwise be missed in X-ray surveys, providing a more complete census of the cluster population and enabling a more accurate assessment of intrinsic scatter in scaling relations.

Overall, our study highlights that X-ray surveys may underestimate the true diversity of cluster X-ray properties. Richness-based optical selection is a powerful complement, revealing massive systems that are faint in X-rays. Future surveys combining optical, X-ray, and SZ observations will be essential to fully characterize the cluster population and to quantify selection biases for both astrophysical and cosmological applications.

\begin{acknowledgements}
SA acknowledges INAF grant  
``Characterizing the newly  discovered clusters of low surface  brightness" and PRIN-MIUR grant
20228B938N ``Mass and selection biases of galaxy clusters: a multi-probe approach", the latter funded by the European 
Union NextGenerationEU, Mission 4 Component 1  CUP C53D2300092 0006. 
This work made use of data supplied by the UK Swift Science Data Centre at the University of Leicester.
This work has been partially supported by the ASI-INAF program I/004/11/4

\end{acknowledgements}

{}

\label{lastpage}

\end{document}